\documentclass[lettersize,journal]{IEEEtran}
\usepackage[flushleft]{threeparttable}
\usepackage{algorithmic}
\usepackage{cite}
\usepackage{array}
\usepackage[caption=false,font=normalsize,labelfont=sf,textfont=sf]{subfig}
\usepackage{etoolbox}
\usepackage{algorithm}
\usepackage[]{footmisc}
\usepackage{textcomp}
\usepackage{enumitem}
\usepackage{stfloats}
\usepackage{url}
\usepackage{verbatim}
\usepackage{graphicx}
\usepackage{multirow}
\usepackage{mathtools}
\usepackage{float} 

\usepackage[flushleft]{threeparttable}
\usepackage{multicol}
\usepackage{array,colortbl,xcolor}
\usepackage{color,soul}
\usepackage{xcolor}
\usepackage{tikz}

\usepackage{mathtools}
\usepackage{amsmath,amsfonts}
\usepackage{amssymb,mathtools}
\newcommand{\E}{\mathbb{E}}
\usepackage{dashbox}%
\usepackage{algorithm}
\begin{document}

\title{ARIA: Anti-diagonal Reduced-cost \\Index-masked Adaptive BLMS Algorithm}

\author{Mohd~Tasleem~Khan,~\IEEEmembership{Senior~Member,~IEEE}}%

\maketitle

\begin{abstract}
The block least-mean-square (BLMS) algorithm supports high-throughput adaptive filtering but incurs high computational cost as the filter length and block size increase. This letter proposes anti-diagonal reduced-cost index-masked (ARIA) adaptive BLMS, which reduces coefficient-update multiplications through anti-diagonal index masking and tap-wise step-size scaling while keeping the filtering exact. Mean-square error (MSE) analysis establishes an explicit step-size bound and the misadjustment. An extension, ARIA-MF (ARIA with masked filtering), also applies the mask during filtering; its analysis shows that the excess MSE is redistributed across block rows while its block average is approximately preserved. Simulations validate the analysis and show very good agreement with the estimates.
\end{abstract}

\begin{IEEEkeywords}
Adaptive filter, BLMS, computational cost, mean-square error, step-size bound.
\end{IEEEkeywords}

\section{Introduction}
\IEEEPARstart{A}{daptive} filters based on the least-mean-square (LMS) algorithm are widely used in system identification and related signal-processing applications~\cite{haykin1996adaptive, sayed2003fundamentals, nascimento2000learning, khan2024application, fellmann2023block}. In conventional LMS, however, the sample-by-sample coefficient update forms a feedback loop that limits the clock rate~\cite{khan2023stochastic}, a bottleneck for high-throughput applications such as 6G~\cite{shamsabadi2025exploring, khan2022asic}. The block LMS (BLMS) algorithm removes this bottleneck by updating the coefficients once per block of $L$ samples, so the $L$ outputs and the gradient estimate can be computed in parallel, with data and coefficients reused across the block~\cite{burrus1971block, clark1981block, lee2012step, lim1997optimum, feuer1985performance}. The convergence behaviour and step-size bounds of BLMS have also been analysed~\cite{feuer1985convergence, lee2012step, khan2023stochastic}.

Block processing remains an active topic in applications with long impulse responses, such as channel estimation, echo cancellation and active noise control, with recent examples including sparsity-enhanced fast block constrained LMS~\cite{zhang2026fbclms}, variable-step-size adaptation~\cite{wu2025regularized}, and convergence analysis of partitioned-block frequency-domain filters~\cite{luo2026convergence}.

However, the parallel computation of the $L$ outputs and the gradient in BLMS comes at a cost. For a filter of length $N$, the coefficient update requires on the order of $LN$ multiplications per block, which dominate area and power as $L$ and $N$ grow~\cite{khan2022asic}. Block-adaptive subband (partitioned) structures face a similar burden~\cite{yang2021bsf, morgan1995delayless}. This multiplication cost is particularly relevant for resource-constrained implementations, where finite word length and quantisation also affect the weights and update terms~\cite{khan2023analyzing}.

Several approaches reduce the complexity of BLMS. Frequency-domain BLMS uses FFTs to lower the per-sample cost for long filters~\cite{lee2012step, yang2021convergence}, but transform overhead and latency limit its benefit for short and moderate filter lengths~\cite{hazarika2023efficient}. Fast exact algorithms~\cite{benesty1990fast, dwivedi2018fast} require more intricate structures. Delayed block adaptation~\cite{khan2023stochastic} relaxes the update dependency, but outdated gradients reduce the stability margin and slow convergence. A complexity-reduction scheme should therefore preserve the stability and misadjustment performance of the original algorithm~\cite{solo1994adaptive}.

This letter takes a different route in the time domain: it keeps the update current and masks a structured subset of its correlation terms, trading a controlled loss in steady-state accuracy for fewer multiplications, and is therefore complementary to existing schemes~\cite{benesty1990fast, dwivedi2018fast}. The proposed anti-diagonal reduced-cost index-masked (ARIA) BLMS algorithm retains only selected correlation products in the update. Because the number of retained terms varies across taps, tap-wise gain scaling preserves the mean convergence behaviour of BLMS. ARIA keeps the filtering exact, while its extension, ARIA-MF (masked filtering), also reduces the filtering multiplications. The main contributions are:
\begin{itemize}
    \item ARIA, an index-masked BLMS framework with exact filtering and fewer update multiplications.
    \item Mean-square error (MSE) analyses of ARIA, yielding its explicit step-size bound (\ref{eq:explicit}), and misadjustment (\ref{eq:misadj}).
    \item ARIA-MF, which also reduces filtering multiplications relative to a BLMS, with row-wise excess MSE analysis.
\end{itemize}
The remainder of this letter is organised as follows: Section~II presents the ARIA framework and analysis; Section~III extends the analysis to the ARIA-MF; Section~IV discusses the results; and Section~V concludes.

\section{ARIA Framework and Analysis}\label{sec:framework}
For block $k$, let $\mathbf{X}_k\in\mathbb{R}^{L\times N}$ be the input data matrix with entries
\begin{equation}
X_k(j,i)=x(kL-j-i+2),
\end{equation}
let $\mathbf{w}_k\in\mathbb{R}^{N}$ be the coefficient vector, $\mathbf d_k\in\mathbb R^{L}$ the plant output block, and $\mathbf z_k$ the additive noise block. BLMS computes
\begin{align}
\mathbf{e}_k &= \mathbf{d}_k+\mathbf z_k-\mathbf{X}_k\mathbf{w}_k, \label{eq:error}\\
\mathbf{w}_{k+1} &= \mathbf{w}_k+\hat\mu\,\mathbf{X}_k^{T}\mathbf{e}_{k}, \label{eq:blms}
\end{align}
where $\hat{\mu}=\mu/L$ is the scaled step size, $\sigma^2$ denotes the input power, and $u=\hat{\mu}\sigma^2$ is the normalised step size.

ARIA reduces the cost of (\ref{eq:blms}) by discarding part of the correlation products in $\mathbf X_k^T\mathbf e_k$. Let $\mathbf{T}$ be the $L\times N$ binary mask with $T(j,i)=1$ if $i\leq N+1-j$ and $T(j,i)=0$ otherwise, so the zeros form a triangle in the lower-right corner. Applying $\mathbf{T}$ with a common step size lets the taps adapt at different rates, since taps with larger index retain fewer correlation terms. To equalise tap-wise convergence, each tap $i$ is assigned an offline-computed compensation factor $g_i$, which is incorporated into its step size as $\hat\mu g_i$ and depends on the number $c_i$ of mask rows retaining that tap:
\begin{equation}
c_i=\min(L,N+1-i),\qquad g_i={L}/{c_i}.
\label{eq:gain}
\end{equation}
With $\mathbf{C}=\operatorname{diag}(c_i)$ and $\mathbf{G}=\operatorname{diag}(g_i)$, it follows that $\mathbf G\mathbf C=L\mathbf I$. Defining $\tilde{\mathbf X}_k=\mathbf T\odot\mathbf X_k$, where $\odot$ is the element-wise product, the proposed update is
\begin{equation}
\mathbf{w}_{k+1}=\mathbf{w}_k+\hat\mu\,
\mathbf{G}\tilde{\mathbf X}_k^T\mathbf{e}_{k},
\label{eq:proposed}
\end{equation}
with the exact error (\ref{eq:error}). Taps $1$ to $N-L+1$ are retained in all $L$ rows and have $g_i=1$, so only the last $L-1$ taps are rescaled. The discarded products are never computed, so no explicit masking operation is required.

Fig.~1 compares the MSE of BLMS, its masked version and ARIA. The input is white Gaussian with $\sigma^2=1$, the unknown system is a unit-norm FIR filter with $N-L+1$ taps, the noise variance is $-60$~dB, and results are averaged over 250 runs. All algorithms use $N=32$, $L=16$ and $\hat\mu=0.005$, which is inside the BLMS stability bound \cite{feuer1985convergence}. In the masked BLMS, row $j$ of each block uses only the first $N+1-j$ taps, so $\mathbf T$ acts on $\mathbf X_k$ in both the filtering and the update. ARIA applies $\mathbf T$ in the update only, with the compensation $\mathbf G$. Without compensation, the last taps retain fewer correlation terms and adapt more slowly, so the masked BLMS converges more slowly than BLMS. Its MSE also oscillates with period $L$, because the rows of a block retain different numbers of taps. ARIA restores the BLMS mean convergence rate through $\mathbf G$ and, with exact filtering, shows no oscillation.

The analysis of (\ref{eq:proposed}) adopts the independence assumptions in \cite{feuer1985convergence, tobias2004leaky, khan2023stochastic}: $x_n$ is zero-mean white Gaussian, $\mathbf z_k$ is zero-mean white noise with variance $\xi_{\min}$, independent of the input, and $\mathbf w_k$ is independent of $\mathbf X_k$. The plant $\mathbf w^o$ has $M\le N$ taps (zero-padded to length $N$), so that $\mathbf d_k=\mathbf X_k\mathbf w^o$ and the minimum MSE is $\xi_{\min}$. With the coefficient error $\mathbf v_k=\mathbf w_k-\mathbf w^o$, the error in (\ref{eq:error}) becomes $\mathbf e_k=\mathbf z_k-\mathbf X_k\mathbf v_k$, and (\ref{eq:proposed}) gives
\begin{equation}
\mathbf v_{k+1}=\mathbf v_k-\hat\mu\,\mathbf G\tilde{\mathbf X}_k^T\mathbf X_k\mathbf v_k+\hat\mu\,\mathbf G\tilde{\mathbf X}_k^T\mathbf z_k .
\label{eq:vrec}
\end{equation}
Let $\mathbf x_j$ denote the $j$th row of $\mathbf X_k$ (as a column vector), $\mathbf T_j=\operatorname{diag}(T(j,1),\ldots,T(j,N))$, and $\mathbf a_j=\mathbf T_j\mathbf x_j$. Then $\tilde{\mathbf X}_k^T\mathbf X_k=\sum_{j=1}^{L}\mathbf a_j\mathbf x_j^T$, with $\mathbf T_j^2=\mathbf T_j$ and $\sum_j\mathbf T_j=\mathbf C$.

\begin{figure}[t]
\centering
\includegraphics[width=0.75\linewidth]{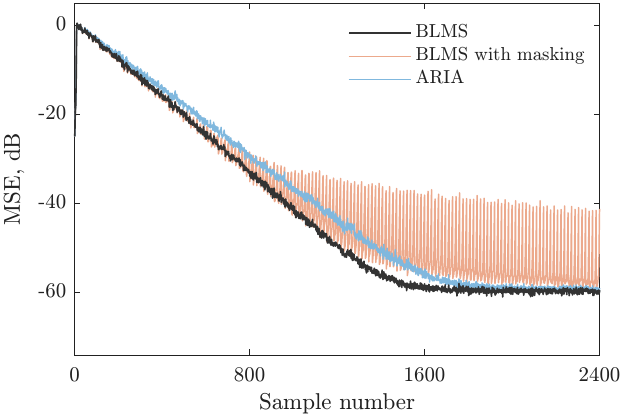}
\caption{MSE learning curves of BLMS and ARIA for $N=32$, $L=16$.}
\label{fig1}
\end{figure}
The mean behaviour follows first. For white input, $\E[\mathbf a_j\mathbf x_j^T]=\sigma^2\mathbf T_j$, hence $\E[\tilde{\mathbf X}_k^T\mathbf X_k]=\sigma^2\mathbf C$, and taking the expectation of (\ref{eq:vrec}) yields
\begin{equation}
\E[\mathbf v_{k+1}]=\left(\mathbf I-\hat\mu\sigma^2\mathbf G\mathbf C\right)\E[\mathbf v_k]=(1-\mu\sigma^2)\,\E[\mathbf v_k].
\label{eq:mean}
\end{equation}
Every tap therefore follows the BLMS mean recursion: since $g_i c_i=L$, the factor $g_i$ exactly compensates for the reduced number of terms. The mean converges if and only if $0<\mu\sigma^2<2$, although the MSE condition derived below is stricter.

For the second moment, let $\mathbf K_k=\E[\mathbf v_k\mathbf v_k^T]$. From (\ref{eq:vrec}), the independence assumptions, $\E[\mathbf z_k]=\mathbf 0$ and $\hat\mu L=\mu$,
\begin{equation}\label{eq:K}
\mathbf K_{k+1}=(1-2\mu\sigma^2)\mathbf K_k+\hat\mu^2\,\mathbf G\,\boldsymbol\Phi(\mathbf K_k)\,\mathbf G +\hat\mu^2L\sigma^2\xi_{\min}\,\mathbf G,
\end{equation}
where $\boldsymbol\Phi(\mathbf K)=\E[\tilde{\mathbf X}_k^T\mathbf X_k\mathbf K\mathbf X_k^T\tilde{\mathbf X}_k]$. The noise term follows from $\E[\tilde{\mathbf X}_k^T\mathbf z_k\mathbf z_k^T\tilde{\mathbf X}_k]=\xi_{\min}\sigma^2\mathbf C$ and $\mathbf G\mathbf C\mathbf G=L\mathbf G$. To evaluate $\boldsymbol\Phi$, the moment-factoring property of a zero-mean white Gaussian vector $\mathbf x$ and a symmetric $\mathbf K$ is used,
\begin{equation}
\E[\mathbf x\mathbf x^T\mathbf K\mathbf x\mathbf x^T]
=\sigma^4\big(2\mathbf K+\operatorname{tr}(\mathbf K)\mathbf I\big).
\end{equation}
In $\boldsymbol\Phi(\mathbf K)=\sum_{j,l}\E[\mathbf a_j\mathbf x_j^T\mathbf K\mathbf x_l\mathbf a_l^T]$, the terms $j=l$ then equal $\sigma^4\big(2\mathbf T_j\mathbf K\mathbf T_j+\operatorname{tr}(\mathbf K)\mathbf T_j\big)$, and, with row independence, the terms $j\neq l$ equal $\sigma^4\mathbf T_j\mathbf K\mathbf T_l$. Using $\sum_{j\ne l}\mathbf T_j\mathbf K\mathbf T_l=\mathbf C\mathbf K\mathbf C-\sum_j\mathbf T_j\mathbf K\mathbf T_j$,
\begin{equation}
\boldsymbol\Phi(\mathbf K)=\sigma^4\Big\{\mathbf C\mathbf K\mathbf C+\sum\nolimits_{j=1}^{L}\mathbf T_j\mathbf K\mathbf T_j+\operatorname{tr}(\mathbf K)\,\mathbf C\Big\}.
\label{eq:Phi}
\end{equation}
In (\ref{eq:K}) and (\ref{eq:Phi}), the diagonal of $\mathbf K_{k+1}$ depends only on the diagonal of $\mathbf K_k$, since $[\mathbf C\mathbf K\mathbf C]_{ii}=c_i^2\kappa_i$, $[\sum_j\mathbf T_j\mathbf K\mathbf T_j]_{ii}=c_i\kappa_i$, and the trace term involves only diagonal entries. Hence the diagonal entries $\kappa_i=[\mathbf K_k]_{ii}$ obey, with $g_ic_i=L$ and $u=\hat\mu\sigma^2$,
\begin{equation}
\kappa_i^{(k+1)}=\rho_i\,\kappa_i^{(k)}+u^2Lg_i\sum\nolimits_{p=1}^{N}\kappa_p^{(k)}+\frac{u^2L\xi_{\min}}{\sigma^2}\,g_i,
\label{eq:krec}
\end{equation}
where $\rho_i=1-2uL+u^2L(L+g_i)$. In vector form,
\begin{equation}
\boldsymbol\kappa^{(k+1)}=\boldsymbol\Gamma\boldsymbol\kappa^{(k)}+\frac{u^2L\xi_{\min}}{\sigma^2}\,\mathbf g,\qquad
\boldsymbol\Gamma=\operatorname{diag}(\rho_i)+u^2L\,\mathbf g\mathbf 1^T,
\label{eq:kvec}
\end{equation}
with $\mathbf g=[g_1,\ldots,g_N]^T$ and initial values $\kappa_i^{(0)}=(w^o_i)^2$ for $\mathbf w_0=\mathbf 0$; for $g_i\equiv1$ this is the BLMS recursion. The learning curve is the a-priori block MSE
\begin{equation}
\xi_k=\xi_{\min}+\sigma^2\sum\nolimits_{i=1}^{N}\kappa_i^{(k)} ,
\label{eq:mse}
\end{equation}
which is the same for all rows of the block because the filtering is exact.

The off-diagonal entries of $\mathbf K_k$ do not affect the diagonal ones and have no forcing term, because the noise term in (\ref{eq:K}) is diagonal. For $i\ne l$, the entry $[\mathbf K_k]_{il}$ is multiplied in each iteration by $1-2uL+u^2g_ig_l\big(c_ic_l+\min(c_i,c_l)\big)$. Since $g_ig_lc_ic_l=L^2$ and $g_ig_l\min(c_i,c_l)=L\min(g_i,g_l)$, this factor equals $1-2uL+u^2L\big(L+\min(g_i,g_l)\big)=\min(\rho_i,\rho_l)\in(0,1)$ whenever $\max_i\rho_i<1$. The off-diagonal entries therefore decay whenever the diagonal ones are bounded, and stability reduces to the spectrum of $\boldsymbol\Gamma$. Since $\boldsymbol\Gamma$ is a rank-one update of a diagonal matrix, its eigenvalues $\beta$ satisfy $\prod_i(\rho_i-\beta)\,q(\beta)=0$ with $q(\beta)=1+u^2L\sum_i g_i/(\rho_i-\beta)$. Because $\rho_i=(1-uL)^2+u^2Lg_i>0$, $\boldsymbol\Gamma$ is non-negative, and $q$ increases on $(\max_i\rho_i,\infty)$ from $-\infty$ to $1$. The largest eigenvalue is thus the unique zero of $q$ above $\max_i\rho_i$, and it is below one if and only if $\max_i\rho_i<1$ and $q(1)>0$. Using $1-\rho_i=uL[2-u(L+g_i)]$ and $g_{\max}=L$, these conditions become
\begin{equation}
u<\frac{1}{L},\qquad
F(u)=\sum\nolimits_{i=1}^{N}\frac{u\,g_i}{2-u\,(L+g_i)}<1 .
\label{eq:cond}
\end{equation}
The first condition is equivalent to $\mu\sigma^2<1$ and is stricter than the mean condition. Since $c_i=L$ for $N-L+1$ taps and $c_i=1,\ldots,L-1$ for the remaining ones,
\begin{equation}
F(u)=\frac{(N-L+1)\,u}{2-(L+1)u}+\sum\nolimits_{c=1}^{L-1}\frac{L\,u}{2c-L(c+1)u}.
\end{equation}
For $u\in(0,1/L)$ every denominator is positive, so $F$ is increasing, with $F(0)=0$ and $F(u)\to\infty$ as $u\to1/L$ (due to the taps with $g_i=L$). The critical step size $u_{\mathrm{crit}}$, defined as the root of $F(u)=1$ in $(0,1/L)$, is therefore unique. An explicit sufficient bound is obtained from $g_{\max}=L$ and $\sum_ig_i=N-L+1+LH_{L-1}$, where $H_n=\sum_{m=1}^{n}1/m$ and $H_0=0$. For $u<1/L$, $2-u(L+g_i)\ge 2-2uL>0$, and hence
\begin{equation}
F(u)\le\frac{u\sum_ig_i}{2-2uL}<1 \ \Longleftarrow\
u<\frac{2}{N+L+1+L\,H_{L-1}}.
\label{eq:explicit}
\end{equation}
For $g_i\equiv 1$, $u_{\mathrm{crit}}$ takes the BLMS form, $2/(N+L+1)$~\cite{feuer1985performance}. The effective filter length increases by $LH_{L-1}$, implying $u<1/L$. The design step $u=1/(N+L+1)$ satisfies (\ref{eq:explicit}) if $LH_{L-1}<N+L+1$. Because the bound concerns the block-rate state $\mathbf K_k$ and every row error is bounded by $\xi_{\min}+\sigma^2\operatorname{tr}(\mathbf K_k)$, it covers all rows of the block. Finally, the steady-state accuracy follows from (\ref{eq:kvec}). With $\mathbf D=\operatorname{diag}(1-\rho_i)$ and the Sherman--Morrison formula \cite{maponi2007solution},
\begin{equation}
\boldsymbol\kappa^{(\infty)}=\frac{u^2L\xi_{\min}}{\sigma^2}(\mathbf I-\boldsymbol\Gamma)^{-1}\mathbf g
=\frac{u^2L\xi_{\min}}{\sigma^2}\,\frac{\mathbf D^{-1}\mathbf g}{1-F(u)},
\end{equation}
where $u^2L\sum_ig_i/(1-\rho_i)=F(u)$. The excess MSE is $\sigma^2\sum_i\kappa_i^{(\infty)}=\xi_{\min}F/(1-F)$, so the misadjustment is
\begin{equation}\label{eq:misadj}
\mathcal M=\frac{\xi_{\mathrm{ex}}}{\xi_{\min}}=\frac{F(u)}{1-F(u)}\approx\frac{u}{2}\,N_g,
\end{equation}
where $N_g=\sum\nolimits_{i=1}^{N}g_i=N-L+1+LH_{L-1}$ and the approximation holds for $u\ll u_{\mathrm{crit}}$. For BLMS, $N_g=N$ and $\mathcal M_{\mathrm{BLMS}}=Nu/[2-(N+L+1)u]$. For small $u$, ARIA has $N_g/N=1+[LH_{L-1}-(L-1)]/N$ times the excess MSE of BLMS, which is small when $L\ll N$. 


\section{Analysis Extension: ARIA-MF}
\label{sec:ext}
The mask can also be applied to the filtering step, $\mathbf{e}_k=\mathbf{d}_k+\mathbf z_k-\tilde{\mathbf X}_k\mathbf{w}_k$, with the update (\ref{eq:proposed}) unchanged. This variant is called ARIA-MF (ARIA with masked filtering). Since the last row retains only $N-L+1$ taps, an $M$-tap plant is modelled without filtering error only if $M\le N-L+1$, which is assumed below.

The analysis of Section~\ref{sec:framework} carries over with a single change. For $M\le N-L+1$, $\tilde{\mathbf X}_k\mathbf w^o=\mathbf d_k$, so $\mathbf e_k=\mathbf z_k-\tilde{\mathbf X}_k\mathbf v_k$ and $\mathbf X_k$ is replaced by $\tilde{\mathbf X}_k$ in the second term of (\ref{eq:vrec}). Since $\E[\mathbf a_j\mathbf a_j^T]=\sigma^2\mathbf T_j$, again $\E[\tilde{\mathbf X}_k^T\tilde{\mathbf X}_k]=\sigma^2\mathbf C$, so the mean recursion (\ref{eq:mean}) is unchanged. In the second moment, only the terms $j=l$ of $\boldsymbol\Phi$ differ: $\operatorname{tr}(\mathbf K)\mathbf T_j$ in (\ref{eq:Phi}) becomes $\operatorname{tr}(\mathbf T_j\mathbf K)\mathbf T_j$. The noise term, $\rho_i$ and the off-diagonal behaviour are therefore unchanged, and since $\min(c_i,c_p)$ rows retain both taps $i$ and $p$, only the coupling term of (\ref{eq:krec}) is replaced,
\begin{equation}
u^2Lg_i\sum\nolimits_{p=1}^{N}\kappa_p^{(k)}\;\longrightarrow\;u^2g_i^2\sum\nolimits_{p=1}^{N}\min(c_i,c_p)\,\kappa_p^{(k)},
\label{eq:krecmf}
\end{equation}
so that $\boldsymbol\Gamma$ in (\ref{eq:kvec}) becomes $\boldsymbol\Gamma_{\mathrm{MF}}$.

The step-size bound is not tightened. Since $g_ic_i=L$, the new coupling coefficient is no larger than $u^2Lg_i$, so $\boldsymbol\Gamma_{\mathrm{MF}}\le\boldsymbol\Gamma$ entry-wise, and by the monotonicity of the spectral radius every step that is stable for ARIA is also stable for ARIA-MF. The explicit bound (\ref{eq:explicit}) is therefore sufficient here as well, and $u<1/L$ remains necessary because $\rho_i$ is unchanged. As in (\ref{eq:cond}), for $u<1/L$ the step is stable if and only if $F_{\mathrm{MF}}(u)<1$, where $F_{\mathrm{MF}}(u)$ is the spectral radius of the non-negative matrix with entries
\begin{equation}
\frac{u\,g_p^{2}\min(c_p,c_q)}{L\,[2-u(L+g_p)]}.
\end{equation}
For ARIA this matrix is rank one and $F_{\mathrm{MF}}$ reduces to $F$ in (\ref{eq:cond}); in general $F_{\mathrm{MF}}\le F$, so the exact threshold is higher than for ARIA. It is unique, since $F_{\mathrm{MF}}$ increases with $u$ and diverges as $u\to1/L$, but it has no scalar form and is found numerically.

The masked filtering also changes how the excess MSE is distributed within a block. In row $j$ the error is $e=z-\mathbf a_j^T\mathbf v$, so the a-priori MSE of that row is $\xi_{\min}+\sigma^2\operatorname{tr}(\mathbf T_j\mathbf K_k)=\xi_{\min}+\sigma^2\sum_{i\le N+1-j}\kappa_i^{(k)}$, in place of (\ref{eq:mse}). For small $u$ the coupling term is of higher order and $\rho_i\approx1-2uL$, so that
\begin{equation}
\kappa_i^{(\infty)}\approx {g_i\,u\,\xi_{\min}}/{2\sigma^2},
\end{equation}
i.e., the steady-state weight-error variance of tap $i$ is $g_i$ times its BLMS value. Relative to the BLMS excess $Nu\xi_{\min}/2$, the excess of row $j$ is therefore $R(j)$, with
\begin{equation}
\underset{j=1,\ldots,L}{R(j)}=\frac{1}{N}\sum_{i=1}^{N+1-j}g_i=\frac{(N-L+1)+L\,(H_{L-1}-H_{j-1})}{N},
\label{eq:profile}
\end{equation}
which decreases from $R(1)=N_g/N$ to $R(L)=(N-L+1)/N$ and averages to $\frac1L\sum_jR(j)=1$, since tap $i$ appears in $c_i$ rows and $g_ic_i=L$. ARIA-MF thus has the same block-averaged excess MSE as BLMS to first order, whereas ARIA has $N_g/N$ times as much, and the excess is redistributed across the rows, producing a ripple of period $L$ in the learning curve.

\begin{figure}[t]
\centering
\includegraphics[width=0.70\linewidth]{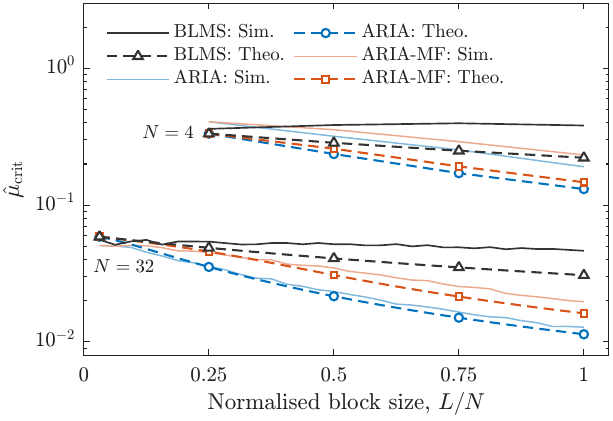}
\caption{Critical step size $u_{\mathrm{crit}}$ versus $L/N$ for BLMS, ARIA and ARIA-MF.}
\label{fig:stability}
\end{figure}
\begin{figure*}[t]
\centering
\includegraphics[width=1\linewidth]{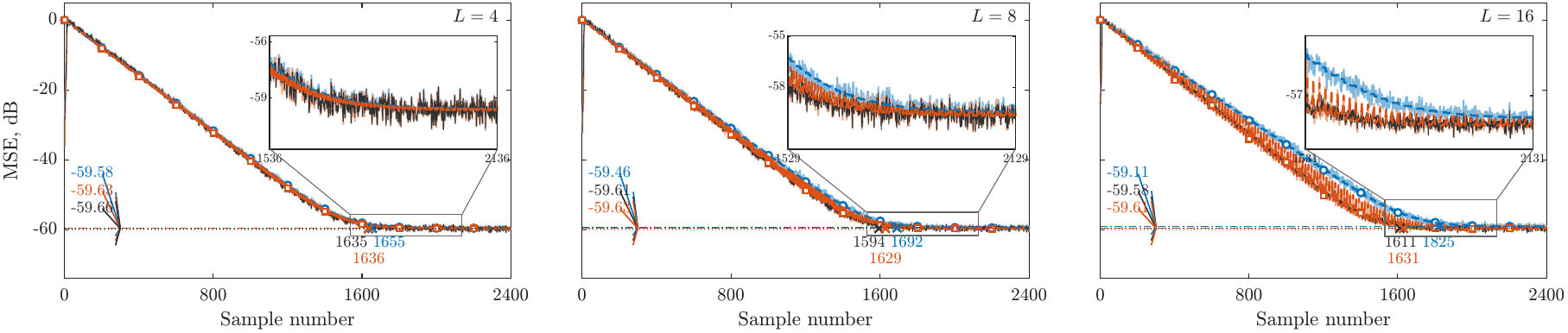}
\caption{MSE learning curves of BLMS (as a baseline), ARIA and ARIA-MF for $N=32$ and $L\in\{4,8,16\}$. Markers: analytical predictions; solid lines: simulations. The horizontal lines indicate the excess error of each model; legends are as in Fig.~2.}
\label{fig:mse1}
\end{figure*}

\section{Results and Discussion}
\label{sec:res}
\subsection{Step-Size Bound}

Fig.~\ref{fig:stability} validates the bound in~\eqref{eq:explicit}
for $N=4,32$ and $L=1,\ldots,N$. For all three algorithms, the critical step size is determined by sweeping $[0.5,1.8]\,u_{\mathrm{crit}}$ in increments of $0.0265\,u_{\mathrm{crit}}$ and identifying the transition from decaying to growing ensemble-averaged error power. Divergence is declared when the steady-state MSE reaches unity.

For $L=1$, all algorithms reduce to LMS and share the same bound \cite{khan2023stochastic},
which decreases approximately as $1/N$. The critical step size
decreases with $L/N$ for all algorithms. Simulated bounds exceed
theoretical ones, as the latter are conservative sufficient
conditions based on the independence assumption. This gap widens
with $L/N$ due to increased dependence between overlapping Hankel
blocks \cite{zhang2020low}. ARIA-MF supports larger step sizes than ARIA in both theory
and simulation. The explicit bound in~\eqref{eq:explicit} is slightly
more conservative than the exact condition in~\eqref{eq:cond};
both remain below the simulated limits. The simulated bounds are
subject to the finite step-size sweep resolution and minor
ensemble fluctuations.
\subsection{Learning Curves}
The simulation conditions are those of Fig.~1, with $N=32$ and $L\in\{4,8,16\}$ ($M=N-L+1$ plant taps). BLMS is simulated only; its theory corresponds to $g_i\equiv1$. Fig.~3 shows that the analytical curves match the simulations in the transient and in the steady-state level.

For $L=4$, the three algorithms are nearly indistinguishable: since
$g_ic_i=L$, all taps share the same mean convergence factor
$(1-\mu\sigma^2)$ in (\ref{eq:mean}), while only $L-1=3$ of the
32 taps have $g_i>1$. Consequently, the mask removes only 6 of
the 128 update products, giving $N_g/N\approx1.14$. The resulting
misadjustment is also similar, with $\mathcal M\approx uN_g/2$
from (\ref{eq:misadj}) yielding about $9\%$ for ARIA versus $8\%$
for BLMS (a difference below $0.05$~dB). Similarly, the ARIA-MF
row profile (\ref{eq:profile}) ranges only from $R(1)\approx1.14$
to $R(L)\approx0.91$, producing a ripple below $0.1$~dB.
As $L$ increases to 8 and 16, the algorithms become more distinct,
with $N_g/N$ rising to approximately 1.43 and 2.19, respectively.

ARIA slows down and settles higher. Its mean rate is unchanged, but the MSE decay is governed by the largest eigenvalue $\beta$ of $\boldsymbol\Gamma$ in (\ref{eq:kvec}), which to first order is $\beta\approx1-2uL+u^2L(L+N_g)$, i.e., the BLMS value with $N$ replaced by $N_g$. The compensated taps, with $g_i$ up to $L$, add a large coupling term $u^2Lg_i\sum_p\kappa_p$, which slows the decay of the common error mode. For $L=16$, this gives $1-\beta\approx0.126$ per block for ARIA against $0.140$ for BLMS. The same $N_g$ enters the misadjustment, so the excess MSE of ARIA is higher by the factor $N_g/N$.

ARIA-MF remains close to BLMS in convergence speed and steady-state error. Its coupling term, $u^2g_i^2\sum_p\min(c_i,c_p)\kappa_p$ in (\ref{eq:krecmf}), is no larger entry-wise than ARIA's, giving $\varrho(\boldsymbol\Gamma_{\mathrm{MF}})\le\varrho(\boldsymbol\Gamma)$. This is because row $j$ contains only the first $N+1-j$ taps, so amplified tail taps contribute to fewer rows. Accordingly, the steady-state excess of row $j$ is $R(j)$ times the BLMS excess. Since the block average of $R(j)$ is one by (\ref{eq:profile}), the block-averaged excess MSE matches BLMS to first order and is lower than ARIA by the factor $N_g/N$.

This comes at the cost of a ripple with period $L$. Since $M=N-L+1$, the plant taps appear in every row and contribute equally, whereas the tail taps appear in fewer rows and are excited through the coupling with plant-tap errors. Their slow decay produces the spikes in Fig.~3, mainly in the transition region. In steady state, the row excess decreases from $R(1)\approx2.19$ to $R(L)\approx0.53$ for $L=16$, a spread of about $0.5$~dB. Both the ripple and spikes increase with $L$.
 \subsection{Computational Complexity}
BLMS requires $2LN$ multiplications per block ($LN$ each for filtering and update). ARIA omits $L(L-1)/2$ update products, a saving of $(L-1)/(4N)$. For $N=32$ and $L=16$, BLMS needs $1024$ multiplications per block and ARIA $904$, a saving of $11.7\%$. ARIA-MF omits a further $L(L-1)/2$ filtering products, giving $2LN-L(L-1)=784$ multiplications ($23.4\%$ saving) and requiring only $N$ input samples per block, against $N+L-1$ for BLMS. In return, an $M$-tap plant requires $N\ge M+L-1$. A BLMS filter sized to the plant ($M=17$) needs only $2LM=544$ multiplications, so ARIA-MF does not outperform a correctly sized BLMS; it is useful when $N$ is fixed or $M$ is unknown.

\section{Conclusion}
This letter proposed ARIA, which reduced the coefficient-update cost of BLMS by anti-diagonal masking with tap-wise step-size scaling, while keeping the filtering exact. ARIA preserved the BLMS mean convergence rate, and its steady-state penalty was small when the block size was small relative to the filter length. ARIA-MF saved further multiplications and redistributed the excess MSE across block rows, with its average approximately preserved. Simulations confirmed the analysis and the trade-off between cost and accuracy.

\bibliographystyle{IEEEtran}
\bibliography{IEEEabrv,ref}
\end{document}